\newcommand{\myorcid}{0000-0003-1851-6411}  
\newcommand{\myname}{Geoff Boeing~\orcidlink{\myorcid} and Yuquan Zhou}

\newcommand{\myaffiliation}{University of Southern California}
\newcommand{\paperdate}{2026}
\newcommand{\papertitle}{Travel Time Prediction from Sparse Open Data}
\newcommand{\papercitation}{Boeing, G. and Y. Zhou. \paperdate. \papertitle. \textit{International Journal of Geographical Information Science}, published online ahead of print. \href{https://doi.org/10.1080/13658816.2026.2628193}{doi:10.1080/13658816.2026.2628193}}
\newcommand{\paperkeywords}{Urban Planning, Transportation Geography, Travel Time, Accessibility, Machine Learning}

\RequirePackage[l2tabu,orthodox]{nag} 
\documentclass[12pt,letterpaper]{article} 

\usepackage[T1]{fontenc}    
\usepackage[utf8]{inputenc} 
\usepackage{ebgaramond}     
\usepackage{tgheros}        

\usepackage[USenglish]{babel} 
\usepackage[strict,autostyle]{csquotes} 
\usepackage[babel=true]{microtype} 

\usepackage{amsmath} 
\usepackage{authblk} 
\usepackage{booktabs} 
\usepackage{caption} 
\usepackage{datetime} 
\usepackage[final]{draftwatermark} 
\usepackage{endnotes} 
\usepackage{geometry} 
\usepackage{graphicx} 
\usepackage{natbib} 
\usepackage{rotating} 
\usepackage{setspace} 
\usepackage{titlesec} 
\usepackage{url} 

\usepackage{hyperref} 
\usepackage{orcidlink} 

\newdateformat{monthyeardate}{\monthname[\THEMONTH] \THEYEAR}

\graphicspath{{.}}

\titleformat{\section}{\normalfont\sffamily\large\bfseries\color{black}}{\thesection.}{0.3em}{}
\titleformat{\subsection}{\normalfont\sffamily\small\bfseries\color{black}}{\thesubsection.}{0.3em}{}
\titleformat{\subsubsection}{\normalfont\sffamily\small\color{black}}{\thesubsubsection.}{0.3em}{}

\hypersetup{
    pdfauthor={\myname},
    pdftitle={\papertitle},
    pdfsubject={\papertitle},
    pdfkeywords={\paperkeywords},
    pdffitwindow=true, 
    breaklinks=true, 
    colorlinks=false, 
    pdfborder={0 0 0} 
}

\begin{document}

\title{\papertitle\footnote{{Citation info: \papercitation}}}
\author[]{\myname}
\affil[]{\myaffiliation}
\date{}

\maketitle

\begin{abstract}

Travel time prediction is central to transport geography and planning's accessibility analyses, sustainable transportation infrastructure provision, and active transportation interventions. However, calculating accurate travel times, especially for driving, requires either extensive technical capacity and bespoke data, or resources like the Google Maps API that quickly become prohibitively expensive to analyze thousands or millions of trips necessary for metropolitan-scale analyses. Such obstacles particularly challenge less-resourced researchers, practitioners, and community advocates. This article argues that a middle-ground is needed to provide reasonably accurate travel time predictions without extensive data or computing requirements. It introduces a free, open-source minimally-congested driving time prediction model with minimal cost, data, and computational requirements. It trains and tests this model using the Los Angeles, California urban area as a case study by calculating naïve travel times from open data then developing a random forest model to predict travel times as a function of those naïve times plus open data on turns and traffic controls. Validation shows that this interpretable machine learning method offers a superior middle-ground technique that balances reasonable accuracy with minimal resource requirements.

\end{abstract}

\section{Introduction}

Travel time prediction underpins many transport planning processes and geographical research problems. It is key to understanding urban accessibility, transport mode and route choice, and location decisions \citep{zhang2015gradient,jenelius2013travel}. Predicting congested travel times (i.e., when traffic congestion impedes flows) requires large volumes of real-time geospatial data on traffic conditions \citep[e.g.,][]{hou2018network}. Such data are proprietary, expensive, and beyond the reach of many planning practitioners and scholars \citep{carrion2012value,wang2011estimating}. This is usually not a problem for well-resourced organizations with highly-skilled labor---however, GIScience technical expertise, complex data and software requirements, and high costs pose insurmountable obstacles for many urban planning researchers, practitioners, and community advocates \citep{giles2022creating}.

Travel time is fundamentally a function of distance and speed, which itself is a function of speed limits, traffic controls, turns, and congestion. Although common \enquote{naïve} travel time prediction methods (which minimize travel distance or street segment traversal time) provide easy solutions, they ignore some of these components and thus usually under-predict real-world driving travel time \citep{ludwig2023traffic,salonen2013modelling,yiannakoulias2013estimating}. However, as we will argue, minimally-congested travel time prediction offers an important yet under-realized middle ground: far more accurate predictions than naïve travel times, yet much lower data and computing requirements than real-time traffic-aware congested travel time prediction.

To address the need for such a middle ground, this article contributes a novel, accurate, easy-to-use, and free method of minimally-congested driving travel time prediction. Our results show a substantial improvement over traditional, naïve methods of travel time prediction common in the urban planning literature. As we show in this article, across the Los Angeles urban area, our average trip's travel time prediction differs from Google's travel time prediction by only 0.34 seconds. In contrast, simple---but common in the literature---naïve travel time prediction minimizing max-speed edge traversal times under-predicts by over 3 minutes on average.

The rest of this article is organized as follows. First it reviews the state of the art in travel time prediction and then considers the less-sophisticated methods that are standard in the current planning literature. Next it describes our proposed model, data sources, and validation techniques. Then it presents our findings, highlighting both the feasibility and quality of our predictions in relation to simple naïve predictions or difficult expensive predictions. Finally, it concludes with a discussion of implications for research and practice.

\section{Background}\label{sec:background}

Modern travel time prediction techniques tend to fall into one of two camps: (1) sophisticated, accurate, but difficult to execute; or (2) naïve, less-accurate, but simple to execute. The former is the state-of-the-art and has been the subject of much GIScience work particularly in the recent computer science and transportation engineering literatures. The latter tends to appear often in urban planning research and practice. Here we summarize the recent advances and current standards in these fields.

\subsection{State-of-the-Art Travel Time Prediction}

Recent studies in the computer science and transportation engineering literatures propose a variety of prediction techniques with heavy computing and data requirements. They usually measure their prediction accuracy by mean absolute percentage error (MAPE), defined in \autoref{eq:mape}:

\begin{equation}
\label{eq:mape}
\text{MAPE} = \frac{100}{n} \sum^{n}_{i=1} \left|\frac{t_i - \hat{t_i}}{t_i}\right|
\end{equation}

where $n$ is the total number of trips, $t_i$ is trip $i$'s observed travel time, and $\hat{t_i}$ is trip $i$'s predicted travel time. Lower MAPEs indicate higher accuracy.  These studies usually yield MAPEs within roughly 10\% of observed driving times. Some researchers use alternative measures of accuracy, such as the percentage of over- or under-prediction \citep{jenelius2013travel}, the absolute mismeasurement in minutes \citep{chiabaut2021traffic}, or degrees of predictability \citep{li2019travel}.

These advanced travel time prediction models require massive data inputs as they are usually trained on millions or billions of disaggregate empirical travel time records. For example, \citet{hou2018network} use 1.5 billion GPS-based travel time records in deep learning models to predict travel time in 13 road segments in St. Louis, yielding a MAPE of 6.8\%. Using more than 100,000 trips from Windows phone GPS data, \citet{woodard2017predicting}'s technique achieves a MAPE of 10.1\% when predicting travel times. \citet{pamula2023estimation} use three months of video sensing data at a 5-minute temporal resolution in a deep learning model to predict origin-destination (OD) matrices in Poland, yielding a 6.8\% MAPE.\@ Other studies use graph neural network models to predict travel times. For example, \citet{wang2023dynamic} use data from 100,000 Chinese ride-hailing trips in graph neural network and recurrent neural network models, yielding a MAPE of 15.4\%. \citet{vankdoth2023deep} achieve low MAPEs using Q-Traffic and TaxiBJ data from Beijing and Chengdu in deep learning and graph neural network models.

Many of these studies focus on predicting travel times only for a specific street segment. For instance, \citet{sharmila2019svm} achieve around 10\% MAPE using GPS data in a support vector machine/particle filter model to predict travel times along specific arterials in Mumbai. \citet{chen2016multi}'s agent-based model uses data from GPS-equipped vehicles at 1-minute intervals across 123 days to achieve a MAPE under 9\% on a 95-mile freeway section in Virginia. Based on empirical travel time from GPS traces, road geometric features, and weather information, \citet{qiu2021machine} predict travel times using decision trees, random forests, extreme gradient boosting, and long short-term memory neural networks on the I-485 freeway in Charlotte, North Carolina, yielding MAPEs between 6--18\%.

As we have seen, these studies train or validate their models with a wide variety of bespoke data sources often collected from GPS devices or proprietary sources. The Google Maps API offers another common source. Though a black-box, many researchers use Google's travel times as the best-available proxy of real-world travel times, given its ubiquity and accuracy and the challenges of otherwise directly obtaining sufficient empirical travel data \citep[e.g.,][]{goudarzi2018travel,stanojevic2019mapreuse,ludwig2023traffic}. Prior studies have validated it as a robust proxy for real-world travel times \citep[e.g.,][]{wu2019comparing,wagner2020travel}.

\subsection{Path Solving in Planning Research}

Central to all of the preceding travel time prediction research is the concept of \textit{path solving}: identifying the shortest path between an origin and destination in a spatial network model. In this case, the shortest path is defined as the path that minimizes travel time and could be realized methodologically in multiple ways. In the planning literature, studies variously minimize Euclidean distance, network distance, network edge traversal time, uncongested travel time, or congested travel time. We discuss their trade-offs in this section.

The actual path solving methods used in urban planning research and practice tend to differ substantially from the state of the art travel time prediction in the transportation engineering and computer science literatures. The preceding section's models require extensive technical capacity, instrumentation, and data. In general, the planning literature's implementations of path solving tend to be simpler to execute, but potentially naïve and less-accurate. For example, many accessibility studies employ merely a (very) rough proxy for travel time by minimizing Euclidean distance traveled instead \citep[e.g.,][]{macfarlane2021modeling,pearsall2020locating}. This offers the substantial benefit of simplicity (no network model, travel time data, or routing algorithms are needed) but likely offers a poor estimate of access in terms of travel time. However, this latter claim is difficult to test as this literature rarely provides validation results or accuracy indicators like MAPE, so there is little basis for direct comparison.

Other studies improve on this by instead solving shortest paths by minimizing network distance traveled. These studies often use open data from OpenStreetMap or the US Census Bureau, or proprietary tools like Esri to measure network distances between origins and destinations \citep[e.g.,][]{mckenzie2020urban, jiao2021measuring, nicoletti2023disadvantaged, logan2019evaluating, tsou2005accessibility}. This offers the benefit of more-realistic distances between origins and destinations, but does not account for travel speeds. Other studies refine this by incorporating speed limit data into the network model to solve shortest paths by traversal time \citep{kuai2017examining,williams2020parks,he2020evaluating,salonen2013modelling,scott2008role,neutens2010equity,wang2013planning}. This approach is similar to---but distinct from---uncongested travel time. The former merely minimizes the sum of the ratio of street segment length and speed limit, whereas the latter accounts for stops, signals, and turns. Many tools (e.g., Esri Network Analyst, OSRM, R5, etc.) can incorporate such impedances, as any routing algorithm should be able to do this. The key challenge, rather, is estimating accurate time penalties from the spatial presence of these traffic control features instead of just assigning penalties heuristically \citep[e.g.,][]{yiannakoulias2013estimating}.

Some researchers sidestep the challenges of aquiring a detailed network model and travel speed data by using secondary traffic analysis zone (TAZ) travel times aggregated by metropolitan planning organizations \citep[e.g.,][]{grengs2010intermetropolitan, shin2020disparities, yan2021toward}. Such TAZ-level data are generated by integrated transport models---deriving from the traditional four-step process \citep{johnston_urban_2004}---that combine trip generation, distribution, assignment, and traffic simulation to predict travel times but have extensive data and software requirements. For example, the Southern California Association of Governments runs its Coordinated Travel–Regional Activity Modeling Platform using the Caliper Corporation's proprietary TransCAD software \citep{scag_transportation_models}. Although such models provide empirically grounded travel time predictions, they are resource-intensive, inaccessible to many independent researchers, and their standard TAZ-level aggregations obfuscate point-to-point travel times.

Another proprietary and costly approach uses commercial data like Esri's Streetmap Premium which incorporates historical traffic data from providers like Here, TomTom, and Increment P \citep[e.g.,][]{lin2021impact}. Although these products offer decent accuracy and granular point-to-point routing, their licensing costs and terms prevent usage by many researchers and practitioners, especially for large-scale projects \citep{delmelle2019travel,fu2023comparative}.

Finally, many planning scholars rely instead on the Google Maps API to obtain accurate OD travel times and shortest paths \citep[e.g.,][]{fielbaum2021assessment,costa2021spatial, swayne2021integrating,hu2020estimating,cuervo2022dynamic,chen2020communities,hwang2024measuring}. As mentioned previously, Google travel times are widely accepted as the best-available proxy of real-world travel times in the literature, but come with drawbacks for planning applications: the algorithms are closed-source and---for large batches of queries, such as for simulating metropolitan-scale trip taking---it can become prohibitively expensive and require many API queries. It is also unsuitable for scenario planning with transportation network changes (e.g., adding a new street or changing traffic controls). Nevertheless, evidence shows Google travel times offer a good proxy for observed real-world travel times \citep{lin2021impact,fu2023comparative,alsobky2020estimating,wang2011estimating,wu2019comparing}.

\subsection{Open Problem}\label{sec:problem}

In summary, planning researchers today predict travel distances and times with a range of relatively simple techniques that minimize Euclidean distance, network distance, or street segment traversal time, or use TAZ-to-TAZ travel times or proprietary data like Google travel times (for relatively few trips to keep costs down). These techniques are generally inexpensive and easy to implement. In contrast, today's state-of-the-art techniques in the computer science and transportation engineering literatures require extensive data collection, private data, or much more complicated algorithms that planners are rarely trained to implement. In other words, it is expensive and challenging for the average planner to reproduce these state-of-the-art methods.

Travel time is essentially a function of distance, speed limits, traffic controls, turns, and traffic congestion. Naïve methods tend to predict it from just the first one or two of these components, whereas resource-intensive state-of-the-art methods predict it from all five. But, with open data, the first four components can now be freely acquired in many parts of the world---only congestion data remains universally proprietary, expensive, and inaccessible. Therefore, could there be a middle-ground between expensive, complicated, real-time predictions and overly-simple naïve predictions? Can we predict travel times from free open data on distance, speed limits, traffic controls, and turns? Such predictions may not reflect real-time congestion (which varies drastically throughout the day and week), but would accurately represent minimally-congested travel times.

This middle-ground method should meet three criteria. First, it should only use free software and free data, avoiding expensive API queries or bespoke GPS data collection. Second, it should be a simple and accessible method for urban planners without advanced technical skills. Third, it should provide a substantial improvement in accuracy and spatial resolution over naïve predictions such as minimizing Euclidean distance, network distance, street edge traversal time, and exsiting TAZ-to-TAZ travel times, and approach the accuracy of the state-of-art models. Meeting these three criteria, a middle ground would be inexpensive, easy-to-use, and reasonably accurate.

\section{Methods}

\begin{figure*}[bt!]
    \centering
    \includegraphics[width=1.0\textwidth]{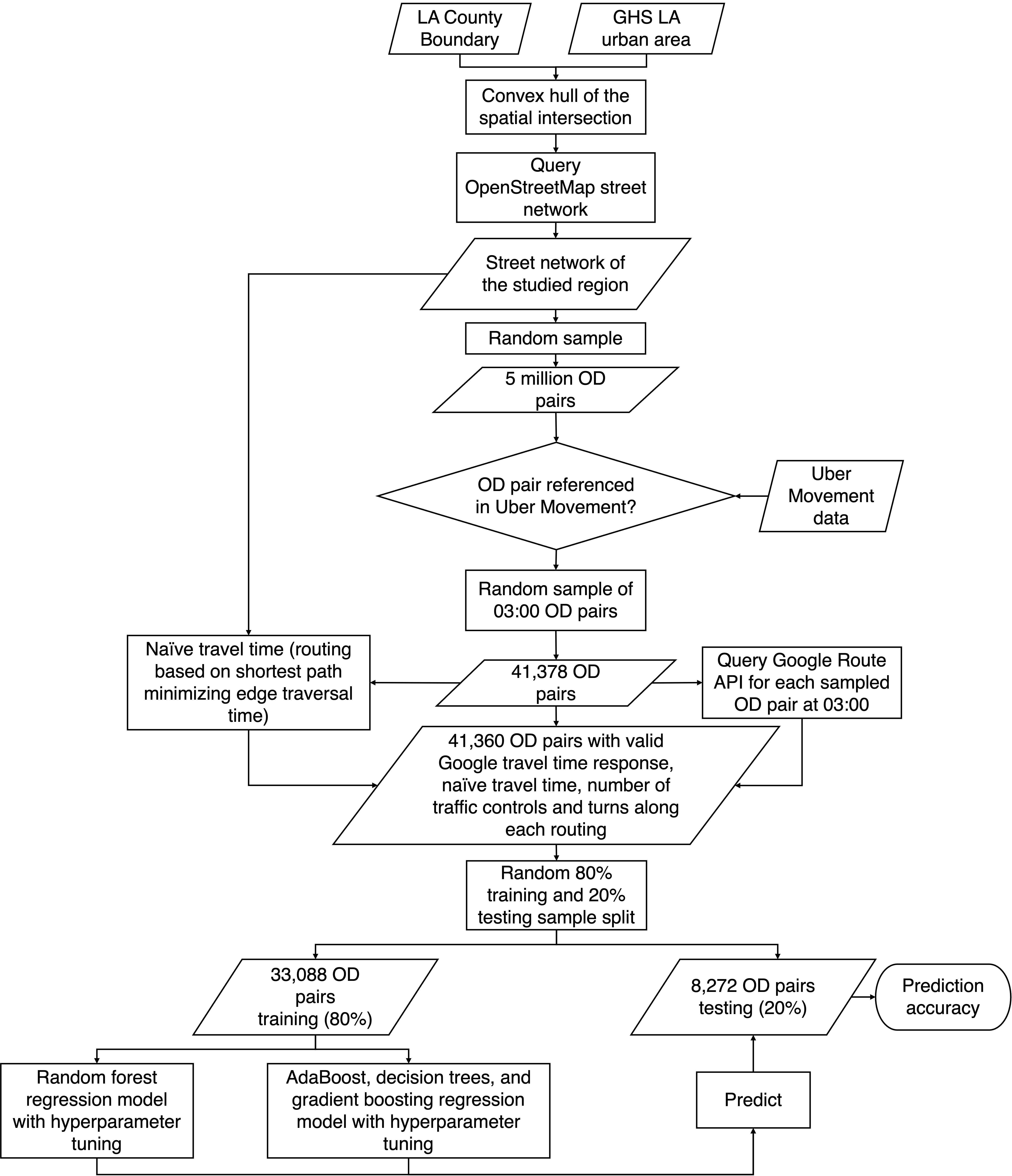}
    \caption{Detailed model workflow.}\label{fig:workflow}
\end{figure*}

This article proposes a novel open-source, reusable, generalizable method to solve this problem. In short, it collects free open data plus a small amount of free data from the Google Maps API to train a local model to predict minimally-congested driving travel times with sparse data and simple computing hardware. \autoref{fig:workflow} shows a detailed workflow.

\begin{table*}[tb!]
    \centering
    \caption{Traffic control elements (as identified by OpenStreetMap tags) present in the graph, demonstrating increased tagging coverage in the 2025 data compared to 2023.}\label{tab:traffic_control_counts}
    \begin{tabular}{lrr}
        \toprule
        Element                        & 2023 count & 2025 count\\
        \midrule
        Crossing                       &  21,560    &  60,742   \\
        Stop sign                      &  26,304    &  40,417   \\
        Traffic signal                 &  15,262    &  19,730   \\
        Mini roundabout                &      44    &      62   \\
        Give way                       &     189    &     491   \\
        \midrule
        Total traffic control elements &  63,359    & 121,442   \\
        Total street intersections     & 127,093    & 127,323   \\
        Total nodes                    & 782,825    & 852,075   \\
        \bottomrule
    \end{tabular}
\end{table*}

\subsection{Input Data}

We define an implementation study area as the convex hull around the intersection of the Los Angeles County boundary and the Los Angeles urban area boundary from the Global Human Settlement Layer's Urban Center Database \citep{florczyk2019description, GHS2019}. This allows us to retain the main urbanized area without adjacent metropolitan areas (such as the Inland Empire). We then model the drivable street network within this study area from OpenStreetMap in November 2023 using the OSMnx package \citep{boeing_modeling_2025} and retaining the strongly connected component, to produce an unsimplified graph with 782,825 nodes and 127,093 identified street intersections. The graph also contains 63,359 tagged traffic control elements. \autoref{tab:traffic_control_counts} summarizes these nodes' tagged traffic controls: these data are sparse and many true intersections lack traffic control information.

To generate OD pairs for training a prediction model, we first over-sample 5,000,000 random node pairs from the street intersections and dead-ends in this graph. We then filter these down to realistic, minimally-congested trip patterns by using the most-recently released Uber movement data \citep{ubermovement2020}. These data derive from Uber trip GPS traces aggregated by tract-to-tract flow and hour of the day. We filter our OD pairs down to those that have a matching real-world trip ($n$ = 1,197,513) that occurred during the 03:00 hour ($n$ = 41,378) to best approximate minimally-congested traffic conditions. These Uber data are not a perfect measure of all real-world trips, but do validate the presence of travel demand between different OD pairs at different times of day.

Then we proxy \enquote{true travel times} by collecting travel times from the Google Maps Routes API.\@ As summarized in \autoref{sec:background}, prior studies have validated Google travel times as a highly accurate proxy for real-world travel times. For each OD pair, the API provides the fastest network path, its travel time, and its length. The API allows users to predict travel times for trips departing in the future: the closer the departure time, the more accurate the prediction. Accordingly, we set the departure time to 03:00 when the traffic was supposed to be minimal and performed the query immediately beforehand (between 02:30--02:40) on 31 January and 1 February 2024. We used Google's \enquote{BEST\_GUESS} traffic-aware model to predict its \enquote{duration\_in\_traffic} travel time. The API was unable to solve 18 OD pairs' routes, resulting in 41,360 OD pairs with a travel time, which we use as the response vector in our prediction model.

\begin{figure*}[tb!]
    \centering
    \includegraphics[width=1.0\textwidth]{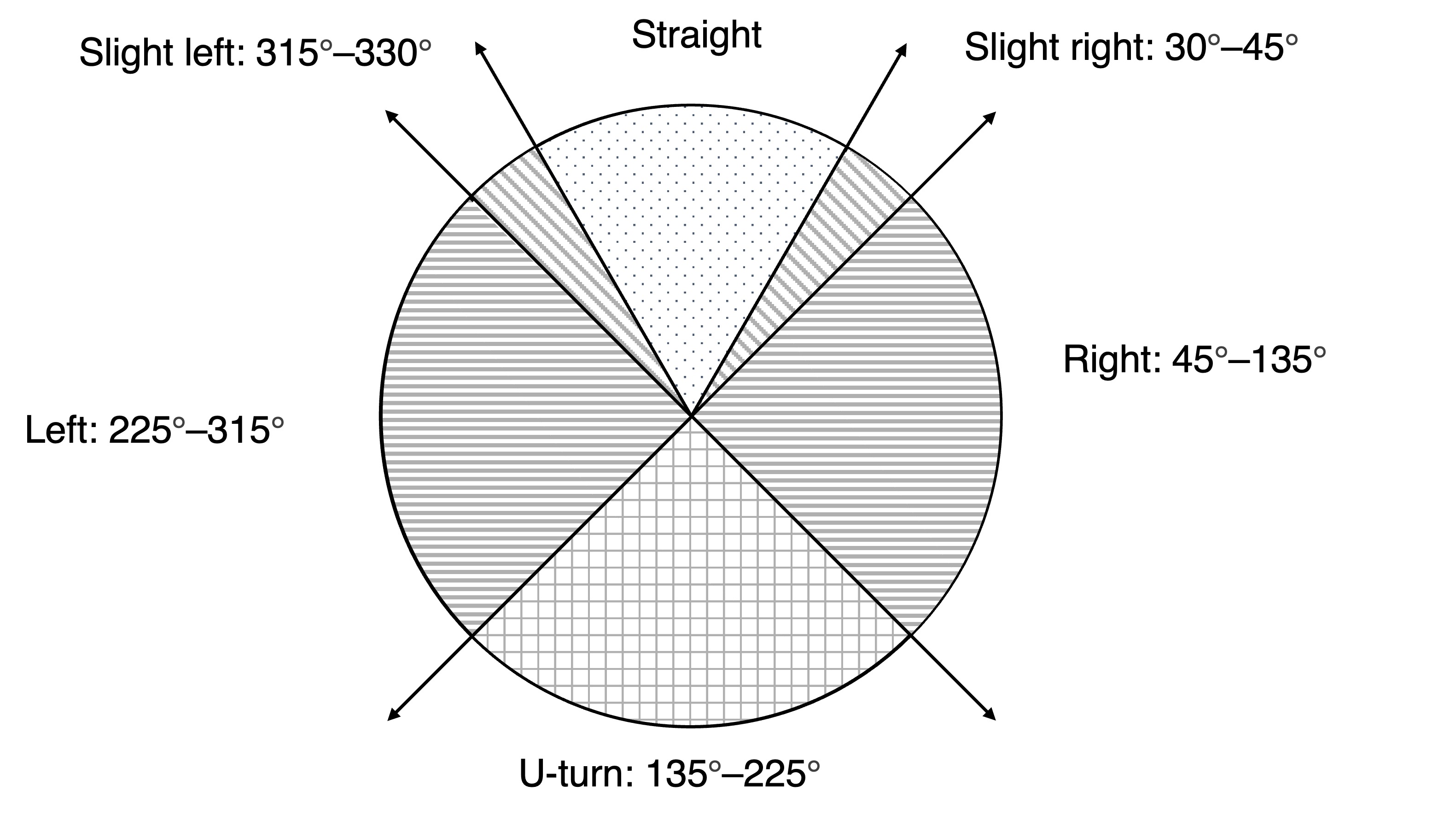}
    \caption{Angular definition of turns in the prediction model.}\label{fig:turns_definition}
\end{figure*}

\subsection{Model Specification and Tuning}

Conceptually, our travel time prediction model has two steps. First, we calculate \enquote{naïve} travel times across a set of trips using open data and open-source software. Second, we train a random forest regression model to predict travel times from our naïve travel times plus a set of covariates.

To calculate the naïve travel times for step one, we solve each OD pair's shortest path using Dijkstra's algorithm to minimize edge traversal time at the speed limit. Then we count the number of traffic control elements across the 5 types in \autoref{tab:traffic_control_counts} (stop signs, traffic signals, pedestrian crossings, give ways, and mini roundabouts), and the number of turns across the 5 types in \autoref{fig:turns_definition} (left, slight left, right, slight right, and u-turn), that were encountered along each OD pair's naïve route. Next we specify a travel time prediction model to predict an OD pair's travel time as a function of its naïve travel time plus these counts of each type of traffic control element. This model's specification is generalized by \autoref{eq:prediction_model}:

\begin{equation}
\label{eq:prediction_model}
y = f(X) + \epsilon
\end{equation}

where $f$ is a prediction model to be trained on our data, $y$ is a length-$n$ response vector (Google travel time), $X$ is a matrix of $n$ observations on 11 predictors (naïve travel time plus counts of the 5 types of traffic control element encountered and counts of the 5 types of turn directions encountered), and $\epsilon$ is random error.

We train candidate models of this specification using several algorithms (decision trees, random forest, gradient boosting, and AdaBoost) chosen to satisfy the three criteria from \autoref{sec:problem} as they have relatively simple data and out-of-the-box computing requirements and handle nonlinearity well. A set of hyperparameters controls this algorithmic training. To tune each's hyperparameters for optimal performance, we split our data into a standard 80/20 training/test split then conduct a randomized grid search with 5-fold cross-validation to minimize the mean absolute error (MAE), defined by \autoref{eq:mae}:

\begin{equation}
\label{eq:mae}
\text{MAE} = \frac{\sum^{n}_{i=1} \left|{y_i - \hat{y_i}}\right|}{n}
\end{equation}

where $n$ is the total trip count, $y_i$ is trip $i$'s Google travel time, and $\hat{y_i}$ is $i$'s predicted travel time.

\subsection{Model Selection and Validation}

Once we have our candidate models tuned and trained, we select our final prediction model and validate the out-of-sample predictions against Google travel times via six accuracy indicators. First, we calculate the MAPE, following \autoref{eq:mape}. Second, we calculate the MAE, following \autoref{eq:mae}. Third, we calculate the mean squared error (MSE), which is more sensitive to outliers than MAE or MAPE, following \autoref{eq:mse} (with the same variables as defined for \autoref{eq:mae}):

\begin{equation}
\label{eq:mse}
\text{MSE} = \frac{\sum^{n}_{i=1} (y_i - \hat{y_i})^2}{n}
\end{equation}

Fourth, we perform difference-in-means ($\delta$) $t$-tests to determine whether our predicted travel times are statistically significantly different (at the 95\% confidence level) from the Google travel times, to detect absolute systematic prediction bias. Despite the large sample size, we expect insignificant $t$-statistics if the predicted travel times' distribution closely matches the Google travel times' distribution. Fifth, we calculate the average pairwise ratio (APR) of our predicted travel times and Google travel times, to detect relative systematic prediction bias, as defined by \autoref{eq:apr} (with the same variables as defined for \autoref{eq:mae}):

\begin{equation}
\label{eq:apr}
\text{APR} = \frac{\sum^{n}_{i=1}\frac{\hat{y_i}}{y_i}}{n}
\end{equation}

Sixth and finally, we calculate the coefficient of determination, $R^2$, to measure our model's ability to explain observed travel time variation.

\subsection{Model Robustness and Interpretability}

As a robustness check, we re-run the model workflow on new OpenStreetMap street network data queried in October 2025, which have much more extensive traffic control coverage than the 2023 data (see \autoref{tab:traffic_control_counts}). To ensure comparability, we use the same set of 40,384 OD pairs (dropping only those with deleted nodes), recalculate naïve travel times on the updated graph, count turns and traffic control features along each path, apply an identical training/testing split, then retrain all candidate models with the same hyperparameter tuning process and the same 2023 Google travel times.

Finally, we conduct a SHapley Additive exPlanations (SHAP) analysis to interpret how the features contribute to travel time predictions. We use the interventional TreeSHAP method for tree-based models, which accounts for non-linearities and interactions among features \citep{lundberg2020local}. This explainer is fitted on the training feature matrix and SHAP values are computed for every OD pair. Each resulting SHAP value represents a feature's contribution (in seconds) to an OD pair's predicted travel time relative to the global mean.

\section{Results}

\subsection{Optimized Hyperparameterization}

After tuning, the four model training algorithms perform similarly across most---but not all---of the accuracy indicators (\autoref{tab:validation_results_2023}). They fall roughly in a similar range for MAPE (7.9--9.0\%), MAE (72.0--80.0 seconds), MSE (12,180--13,570 square seconds), APR (0.99--1.01), and $R^2$ (0.93), demonstrating the robustness of our model to algorithmic particulars. However, the random forest and decision tree models exhibit much lower absolute systematic prediction bias ($\delta$ = 0.34 and 0.13 seconds, respectively) than the gradient boosting and AdaBoost models do (-19.3 and -9.8 seconds, respectively). In other words, all these models have comparable mean errors---but the average random forest and decision tree predictions are statistically insignificantly different from the average Google travel time, whereas gradient boosting and AdaBoost significantly under-predict it. Therefore, we select the random forest model as our \enquote{final} travel time prediction model because it predicts better overall (than the decision tree model) and better avoids systematic under-prediction (than the gradient boosting and AdaBoost models).

Random forests handle nonlinearity well but have much lighter data needs than deep learning models, which satisfies our selection criteria. This final model trains an ensemble of decision trees to predict unobserved response values by essentially dividing the data into boxes then making predictions based on the means of those boxes. Its optimized hyperparameterization uses 400 decision trees, random sampling with replacement in each decision tree, a maximum decision tree depth of 10, use of all available input explanatory features at each split, a requirement of at least two samples to split at a decision node, and the default settings that require at least one sample at a leaf node and apply equal sample weighting. Finally, we check this final model for overfitting by conducting another 5-fold cross-validation across the whole sample using the tuned hyperparameters. The resulting five MAE values (75.4, 74.5, 73.7, 73.2, 74.7) are all very similar, indicating that the model is not overfitted and its hyperparameters are not overly specific to the training set.

\subsection{Model Performance and Validation}

\begin{table*}[tb!]
    \centering\small
    \caption{Out-of-sample prediction accuracy of the 2023 model (chosen random forest model in bold) versus the initial naïve travel time model and the discarded alternative models, all validated against the corresponding 2023 Google travel times. $n$ = 8,272 for each. The six accuracy indicators are (1) MAPE \%, (2) MAE in seconds, (3) MSE in square seconds, (4) difference-in-means ($\delta$) in seconds and its corresponding $t$-test's $p$-value, (5) APR, and (6) $R^2$. See methods for mathematical definitions.}\label{tab:validation_results_2023}
    \tabularnums{\small
    \begin{tabular}{lrrrrrrr}
        \toprule
        Model               & MAPE  & MAE    & MSE      & $\delta$ & $p$  & APR  & $R^2$ \\
        \midrule
        Initial naïve       & 21.15 & 183.68 & 48214.06 &  -182.85 & <0.01 & 0.79 & 0.74  \\
        \textbf{Random forest}       &  \textbf{8.42} &  \textbf{75.34} & \textbf{12181.78} &     \textbf{0.34} &  \textbf{0.78} & \textbf{1.01} & \textbf{0.93}  \\
        Gradient boosting   &  7.86 &  71.95 & 12596.01 &   -19.33 & <0.01 & 0.99 & 0.93  \\
        Decision trees      &  9.00 &  80.04 & 13570.17 &     0.13 &  0.92 & 1.01 & 0.93  \\
        AdaBoost            &  8.20 &  74.04 & 12424.14 &    -9.76 & <0.01 & 1.00 & 0.93  \\
        \bottomrule
    \end{tabular}}
\end{table*}

Compared to the initial naïve travel time calculation, our travel time prediction model exhibits substantial out-of-sample improvement (\autoref{tab:validation_results_2023}). Our model produces a MAPE of 8.4\%, in line with the \textasciitilde10\% MAPEs seen in the state-of-the-art travel time prediction literature, but without their extensive and expensive input data requirements. In comparison, our initial naïve travel time calculation produces a much worse MAPE of 21.2\%. Whereas the naïve travel time's MAE is 183.7 seconds, our model's predicted travel times' MAE is just 75.3 seconds---an improvement by a factor of 2.4. Similarly for MSE, our model improved by a factor of 4.0.

The $t$-test reveals significant differences between the initial naïve travel time calculations and the Google travel times: the $\delta$ of -182.9 seconds corresponds to $p$ < 0.01. In other words, the naïve travel time under-predicts Google travel time by over 3 minutes on average. However, the $t$-test reveals insignificant differences between our model's out-of-sample predictions and the Google travel times: the $\delta$ of 0.34 seconds corresponds to a $p$ value of 0.78. That is, our prediction model over-predicts by less than half a second on average, which is statistically insignificantly different from zero. If we reconsider the aforementioned MAEs in light of these differences-in-means, we can see that the initial naïve travel time calculations significantly and systematically under-predict travel time, but our prediction model shows no such directional bias: its random error averages out between (smaller absolute) over-and under-prediction.

The APR offers another lens on this finding. On average across the OD pairs, the initial naïve travel time calculation under-predicts Google travel time by 21\%, but our prediction model over-predicts it by just 1\%. Our model also explains more of the variation in travel time: its $R^2$ of 0.93 is substantially higher than the 0.74 $R^2$ of the initial naïve travel time calculation. In sum, each of our six accuracy indicators demonstrates that our prediction model drastically improves on naïve travel time calculations---without needing extensive, expensive input data or advanced deep learning software---and offers much more accurate travel time predictions.

\subsection{Model Robustness and Interpretation}

\autoref{tab:validation_results_2025} shows that our modeling process is robust to changes in the underlying data. Compared with the 2023 models (\autoref{tab:validation_results_2023}), the 2025 models achieve similarly high performance across all accuracy indicators, despite extensive changes to the underlying street network model and traffic control elements' coverage (see \autoref{tab:traffic_control_counts}). As with the 2023 models, the mean predicted travel times from the 2025 random forest and decision tree models are not statistically different from the mean Google travel times, whereas other models significantly under-predict. Accordingly, the random forest remains our preferred model because it predicts better than the decision tree model and avoids systematic under-prediction in the gradient boosting and AdaBoost models. For the 2025 models, we re-ran the hyperparameter grid search and identified the same optimal hyperparameters as for the 2023 random forest model. The five cross-validation MAEs (72.2, 71.8, 72.0, 73.2, 72.6) are very similar, again indicating that our model is not overfitted.

\begin{table*}[tb!]
    \centering\small
    \caption{Out-of-sample prediction accuracy of the 2025 model (chosen random forest model in bold) versus the initial naïve travel time model and the discarded alternative models. $n$ = 8,073 for each. Accuracy indicators are the same as defined in \autoref{tab:validation_results_2023}.}\label{tab:validation_results_2025}
    \tabularnums{\small
    \begin{tabular}{lrrrrrrr}
        \toprule
        Model               & MAPE  & MAE    & MSE      & $\delta$ & $p$  & APR  & $R^2$ \\
        \midrule
        Initial naïve       & 21.72 & 188.52 & 49573.15 &  -187.92 & <0.01 & 0.78 & 0.73  \\
        \textbf{Random forest}       &  \textbf{8.12} &  \textbf{73.17} & \textbf{11780.48} &     \textbf{0.15} &  \textbf{0.90} & \textbf{1.01} & \textbf{0.94}  \\
        Gradient boosting   &  7.50 &  69.53 & 12047.36 &   -18.85 & <0.01 & 0.99 & 0.93  \\
        Decision trees      &  8.51 &  76.91 & 12911.55 &    -0.16 &  0.90 & 1.01 & 0.93  \\
        AdaBoost            &  7.86 &  71.70 & 11931.78 &    -9.45 & <0.01 & 1.00 & 0.93  \\
        \bottomrule
    \end{tabular}}
\end{table*}


As expected, naïve travel time is the dominant predictor: its mean absolute SHAP value is 328.1 seconds. Pedestrian crossings contribute an additional 20.2 seconds on average. Among remaining features, left turns have the largest mean absolute SHAP value (5.1 seconds), followed by traffic signals (4.9 seconds), right turns (3.7 seconds), and stop signs (2.8 seconds). Give-way signs and mini-roundabouts have small mean absolute SHAP values, suggesting their effects are trivial. Aside from magnitude, the signs of the SHAP values indicate that more crossings, turns, and signals tend to increase predicted travel time, which is consistent with transportation theory and practice.


\section{Discussion}

Travel time is fundamentally a function of distance, speed limits, traffic controls, turns, and congestion. Naïve methods predict only from the first one or two of these components. Resource-intensive state-of-the-art methods predict it from all five. The planning literature shows that planners often use driving travel time prediction techniques far below the state-of-the-art due to cost, data, and technical capacity constraints. In this study, we propose that a better middle-ground travel time prediction method is feasible and identify three criteria for success: (1) it should use free software and data and avoid bespoke sensor data collection; (2) it should be easy and accessible to use; (3) it should offer better accuracy than naïve predictions (e.g., minimizing Euclidean distance, network distance, street edge traversal time) and better spatial resolution than zone-to-zone travel times.

Our results show that our interpretable machine learning method satisfies these criteria to offer a middle-ground contribution. First, it relies only on open-source software, OpenStreetMap data, and a small free amount of Google travel time training data. Second, our model is easy to use: it does not require a complex computing environment suited for deep learning's extensive data and processing requirements. Rather, our model uses out-of-the-box tools that can be set up in seconds on a standard consumer-grade computer by anyone familiar with Python, the world's most popular programming language. Third, our model is accurate: its MAPE (8.4\%) is much lower than that of the naïve model and in line with state-of-the-art models (\textasciitilde3--17\%) with extensive data and computing requirements.

This middle ground is not necessarily for computer scientists, high-end labs with top-of-the-line computing hardware, or well-resourced agencies with extensive data budgets. Rather, it empowers the scholars and practitioners who need GIScience solutions the most---those working on the frontlines of urban planning and policymaking. The literature shows that these scholars and practitioners often fall back on simple but inaccurate routing models---such as minimizing Euclidean distance, network distance, or edge traversal time---when they lack the technical expertise, computing equipment, extensive data, or funding needed to use the literature's cutting-edge models or commercial solutions. However, this study demonstrates how these naïve methods systematically under-predict real-world travel times. Our model addresses this problem by incorporating sparse open data on traffic control elements and turns into the prediction to achieve an accuracy in line with state-of-the-art methods---but without their data needs or computing requirements. In other words, our model does not replace the state-of-the-art congested-traffic methods for those with extensive resources available to implement them, but rather offers far more accurate predictions for urban planners who would otherwise fall back on simpler naïve models. Inaccurate travel times skew empirical understandings and mislead planning interventions. Our model represents a middle-ground for urban planning scholarship and evidence-informed practitioner interventions.

As demonstrated, this model supports high-resolution point-to-point travel time predictions across a large metropolitan area. Our implementation focuses on Los Angeles but is not inherently tied to it, and future work should emphasize its generalizability in cities with potentially sparser or lower-quality open data. As \citet[p. 230]{breiman2001statistical} notes, \enquote{Algorithmic models accurate in one context must be modified to stay accurate in others. This does not necessarily imply that the way the model is constructed needs to be altered, but that data gathered in the new context should be used in the construction.} Future research should also expand the kinds of training and validation data. Although used by billions of people around the world as the public's \enquote{best available} source of travel time prediction, Google travel time remains a black box derived from users' GPS data. Future work should use other empirical travel time data to further train and validate the model.

We trained this model as a proof-of-concept for minimally-congested times of day, but the same concept can be extended with similar training data for any other time of day if congestion data exist. This is, of course, a big \enquote{if}---and the whole reason why minimally-congested predictions are valuable in the first place. Our goal with this flexible, lightweight model is to improve the prediction accuracy of baseline travel time, bringing naïvely modeled travel time closer to real-world empirical travel time. Practical use cases include scenario planning (such as how hypothetical network or traffic control changes affect travel time), modeling congestion-free baselines to measure the impact of different congestion levels on accessibility, generalized or daily-average accessibility analyses where consistency matters more than overfitting the heterogeneity of specific link-level congestion, and low-resource urban planning practice without proprietary tools, expensive licenses, and intensive data/computing requirements.

\section{Conclusion}

Travel time prediction is central to questions of urban accessibility, mode and route choice, and individuals' location decisions. True congested travel times vary drastically throughout the day and require large volumes of real-time or proprietary geospatial data, as well as complex algorithms, to model. These technical challenges and costs present a hurdle for many urban planners, who often turn to simpler models with lower and often free data and computing requirements. These traditional naïve methods may be easy to implement, but they produce wildly inaccurate predictions.

This article introduced a better middle-ground method to simply but accurately predict minimally-congested driving travel time using free data. It improves on traditional, common, naïve predictions by training on a small one-off collection of free but proprietary high-quality Google data on travel times. Also of note is the sparseness of our open data: OpenStreetMap contains only sporadic information on traffic controls' presence. Even so, our prediction model demonstrates very high accuracy. This offers planning practitioners and scholars a better method to predict travel times for free with much better accuracy than traditional naïve methods offered.

\section*{Data Availability Statement}

The code and data that support the findings of this study are available at \url{https://github.com/gboeing/travel-time-prediction}, minus one training data file which Google's terms of use forbid publicly redistributing. For replication purposes, data equivalent to those in that file are freely available (up to a monthly usage quota) directly from the Google Routes API (\url{https://developers.google.com/maps/documentation/routes}) by signing up for an API key from the Google Cloud Console (\url{https://console.cloud.google.com/}).

\section*{Conflict of Interest Statement}

The authors have no relevant financial or non-financial competing interests to report.

\section*{Acknowledgments}

The authors wish to thank Youngseo Kweon and Jaehyun Ha for additional research assistance.

\section*{Notes on Authors}

\textbf{Geoff Boeing} is an Associate Professor in the University of Southern California's Department of Urban Planning and Spatial Analysis, the Director of USC's Urban Data Lab, and a Nonresident Senior Fellow at the Brookings Institution. His research explores the spatial outcomes of urban planning through network analysis, geospatial data science, and machine learning. He was responsible for study conception, funding, supervision, writing, and revising.
\newline\newline
\noindent\textbf{Yuquan (Wendy) Zhou} is a PhD Candidate in Urban Planning and Development at the University of Southern California. Her research combines geospatial analytics and machine learning to support data-driven urban planning. She was responsible for data curation, analysis, coding, writing, and revising.

\IfFileExists{\jobname.ent}{\theendnotes}{}

\setlength{\bibsep}{0.00cm plus 0.05cm} 
\bibliographystyle{apalike}
\bibliography{references}

\end{document}